\documentclass[prd,twocolumn,showpacs,amsmath,amssymb,aps,showkeys,floatfix,nofootinbib,superscriptaddress]{revtex4}
\pdfoutput=1
\usepackage{graphicx}

\usepackage{nicefrac}

\usepackage[T1]{fontenc}
\usepackage{lmodern}
\usepackage{bm}
\usepackage{verbatim}
\usepackage{enumerate}
\usepackage{amsmath}
\usepackage{latexsym}
\usepackage{mathrsfs}
\usepackage{hyperref}
\usepackage{float}

\def\dd{{\rm d}}
\def\d{{\rm d}}
\def\be{\begin{equation}}
\def\ee{\end{equation}}
\def\bea{\begin{eqnarray}}
\def\eea{\end{eqnarray}}
\def\nn{\nonumber}

\def\rmd{{\rm d}}

\def\PP{\mathcal{P}}
\def\EE{\mathcal{E}}
\def\BB{\mathcal{B}}

\def\nn{\nonumber}

\begin{document}

\title{Internal Robustness of Growth Rate data}

\author{Bryan Sagredo}
\email{bryan.sagredo@ing.uchile.cl}
\affiliation{Cosmology and Theoretical Astrophysics group, Departamento de F\'isica, FCFM, Universidad de Chile, Blanco Encalada 2008, Santiago, Chile}

\author{Savvas Nesseris}
\email{savvas.nesseris@csic.es}
\affiliation{Instituto de F\'isica Te\'orica UAM-CSIC, Universidad Auton\'oma de Madrid, Cantoblanco, 28049 Madrid, Spain}

\author{Domenico Sapone}
\email{dsapone@ing.uchile.cl}
\affiliation{Cosmology and Theoretical Astrophysics group, Departamento de F\'isica, FCFM, Universidad de Chile, Blanco Encalada 2008, Santiago, Chile}

\date{\today}

\begin{abstract}
We perform an \emph{Internal Robustness} analysis (iR) to a compilation of the most recent $f\sigma_8(z)$ data, using the framework of Ref.~\cite{Amendola:2012wc}. The method analyzes combinations of subsets in the data set in a Bayesian model comparison way, potentially finding outliers, subsets of data affected by systematics or new physics. In order to validate our analysis and assess its sensitivity we performed several cross-checks, for example by removing some of the data or by adding artificially contaminated points, while we also generated mock data sets in order to estimate confidence regions of the iR. Applying this methodology, we found no anomalous behavior in the $f\sigma_8(z)$ data set, thus validating its internal robustness.
\end{abstract}


\maketitle

\section{Introduction}\label{intro}
During the last twenty years a plethora of observations suggests that the Universe is undergoing a phase of accelerated expansion at late times. In order to explain this phenomenon, the concept of attractive gravity had to be revised either by introducing a new form of matter  dubbed dark energy, see Ref.~\cite{Sapone:2010iz} for a review, or by altering explicitly the laws of gravity \cite{Tsujikawa:2010zza}. However, the simplest way to account for a phase of accelerated expansion of the Universe within the framework of Friedmann-Lema\^itre-Robertson-Walker (FLRW) cosmology is to simply introduce a cosmological constant ($\Lambda$). While this model gives rise to severe coincidence and fine-tuning problems, current cosmological observations are still compatible with a Universe that is filled by a dark energy component that has the same characteristics of the cosmological constant \cite{Ade:2015xua}.

Nonetheless, these cosmological observations are not accurate enough at the moment to either constrain any potential time evolution of the cosmological constant, which might lead dark energy to cluster, or any modifications of gravity. Despite the fact that the two aforementioned classes of theories can in principle be arbitrarily similar \cite{Kunz:2006wc,Nesseris:2006er}, it is still necessary to be able to discriminate between the currently available models.

Future surveys such as Euclid \cite{Amendola:2012ys}, LSST \cite{Abell:2009aa} and DESI \cite{Aghamousa:2016zmz}, all of which will gather orders of magnitude more data than current surveys, it would be interesting to constrain the dynamical features of gravity and test our assumptions. One way to do this is via the growth of matter density perturbations $\delta_m=\delta \rho_m/\rho_m$ and its logarithmic derivative the growth rate $f=\dd \ln \delta_m/\dd \ln a$. In practice, most of the growth rate measurements are made via the peculiar velocities obtained from Redshift Space Distortions (RSD) measurements \cite{Kaiser:1987qv} coming from galaxy redshift surveys. In general, these surveys measure the perturbations of the galaxy density $\delta_g$, which can be related to (dark) matter perturbations through the bias $b$ via $\delta_g=b\; \delta_m$. Thus, initial growth rate measurements measured the growth rate $f$ divided by the bias factor $b$ leading to the parameter $\beta\equiv f/b$. This parameter is very sensitive to the value of the bias which can vary in the range $b\in[1,3]$, thus making difficult to combine $\beta$ from different surveys and as a result leading to unreliable data sets of $\beta(z_i)$.

As a result, a more reliable parameter was sought and this was found in the combination $f(z)\sigma_8(z)\equiv f\sigma_8(z)$, which can be shown to be independent of the bias \cite{Song:2008qt}, and can be measured either via weak lensing or RSD observations. Still, the current measurements of $f\sigma_8(z)$ (presented in later sections) come from a plethora of different surveys with different assumptions and systematics, thus an approach to study the statistical properties and robustness of the data is imperative.

One such approach would be to perform a tomographic analysis, as was done in Ref.~\cite{Kazantzidis:2018rnb}, where growth data from different years and different redshifts were split into subsamples. Then, it was found that for the more recent data the agreement with the Planck 15 best-fit $\Lambda$CDM cosmology is much improved and is well within $1\sigma$, due to the fact that newer data are at higher redshifts and with higher errors. This approach can clearly highlight inconsistencies in the data, but care must be taken to avoid double-counting of the data.

Another approach to test the internal consistency of low redshift probes was developed in Ref.~\cite{Efstathiou:2017rgv}, where the KiDs data were examined for internal tensions, by performing cuts of the data and examining the cross-correlation measurements of the correlation function that was presented in four tomographic redshift bins by the  KiDs collaboration. Then it was found that the KiDs data might have internal tensions of $\sim 2.2-3.5 \sigma$ significance.

In our paper we choose to follow a more direct approach which is based on Bayesian analysis called ``Internal Robustness'' method and was pioneered in Ref.~\cite{Amendola:2012wc}, see also Refs.~\cite{Heneka:2013hka,Heneka:2014ica} for applications of the method in supernovae data. This is a fully Bayesian approach which is  not only sensitive to the local minimum like a standard $\chi^2$ comparison, but also to the entire likelihood and can in principle detect the presence of systematics in the data set. The main goal of this approach is to identify systematic-contaminated data-points, which can then be further analyzed and potentially excluded if they cannot be corrected.

In this paper we present an application of the ``Internal Robustness'' approach to the currently available growth-rate data in the form of $f\sigma_8(z)$ with the aim to examine the data set for systematics and outlier points in a fully automated manner. The reason we specifically use these data is twofold: first, the growth data are dynamic probes that can in principle discriminate between modifications of general relativity \cite{Nesseris:2007pa}; second, currently there is a tension between low redshift probes, such as the growth data, and the Planck 15 best fit cosmology. Therefore, it is imperative to have a data set whose statistical properties have been demonstrated to be robust before  inferences about modified gravity models are made or the tension with Planck could be explained.

The layout of our manuscript is as follows: in Sec.~\ref{basic-eqs} we provide the basic elements of FLRW cosmology related to our models, in Sec.~\ref{formalism} we briefly review the ``Internal Robustness'' method and its application to the growth data, while in Sec.~\ref{data} we provide the compilation of growth data used in our analysis and finally, we discuss our results in Sec.~\ref{results}.

\section{Basic equations}\label{basic-eqs}

In this section we present the basic equations required in our analysis. We begin with the Hubble parameter in a flat $\Lambda$CDM universe (with a constant equation of state parameter for dark energy $w=-1$), given by
\begin{equation}\label{eq:hubble-parameter}
	H(a)^2=H_0^2 \left[\Omega_{m,0}a^{-3}+(1-\Omega_{m,0})\right] \;,
\end{equation}
where $H_0$ is the Hubble constant, and $\Omega_{m,0}$ is the present day value of the matter density parameter and $a$ is the scale factor.
The matter density can then also be expressed as a function of the scale factor:
\begin{equation}\label{matter_density}
	\Omega_m(a)=\frac{\Omega_{m,0}a^{-3}}{H(a)^2/H^2_0}\;.
\end{equation}
Under the assumption of flat $\Lambda$CDM model, the angular diameter distance takes an analytical expression, given by:
\begin{multline}
\frac{H_0}{c}\frac{\sqrt{\Omega_{m,0}}}{2a} d_A(a)= \, _2F_1\left[\frac{1}{2}\;,\frac{1}{6}\;;1+\frac{1}{6}\;;\left(1-\frac{1}{\Omega_{m,0}}\right)\right]\\
-\sqrt{a}\cdot \, _2F_1\left[\frac{1}{2}\;,\frac{1}{6}\;;1+\frac{1}{6}\;;\left(1-\frac{1}{\Omega_{m}(a)}\right)\right]\;,
\end{multline}
where $_2F_1$ is the hypergeometric function. The matter density perturbations in Fourier space $\delta_m(a,k)$ depend on the underlying cosmological model; for the $\Lambda$CDM scenario, the linear matter perturbations grow according to
\begin{equation}
\delta''_m(a)+\left(\frac{3}{a}+\frac{H'(a)}{H(a)}\right)\delta'_m(a)-\frac{3}{2}\frac{\Omega_m(a)}{a^2}\delta_m(a)=0 \;.
\end{equation}
The equation above has an analytical solution for the growing mode, given by \cite{Silveira:1994yq,Percival:2005vm,Belloso:2011ms}
\begin{equation}
\delta_m(a)= a \cdot \, _2F_1\left(\frac{1}{3}\;,1\;;\frac{11}{6}\;;a^3(1-\frac{1}{\Omega_{m,0}})\right) \;.
\end{equation}
Note that the dependence on the wave number $k$ disappears because of the assumption of small scales approximation.

We define the growth rate $f$ and the root mean square (RMS) normalization of the matter power spectrum $\sigma_8$ as:
\bea
f(a)&=&\frac{\d\log{\delta_m}}{\d\log{a}},\\
\sigma_8(a)&=&\sigma_{8,0}\frac{\delta_m(a)}{\delta_m(1)}\;,\\
\sigma_{8,0}^2&=&\langle \delta(x)^2 \rangle\nn\\
&=& \frac{1}{2\pi^2} \int_0^\infty P(k) W_R^2(k) k^2 dk,
\eea
where $W_R(k)$ is the Fourier transform of a top-hat window function. As already mentioned in Sec.~\ref{intro}, a more robust and reliable quantity that is measured by redshift surveys is the combination of the growth rate $f(a)$ and the RMS $\sigma_8(a)$:
\begin{equation}\label{eq:fs8-parameter}
f\sigma_8(a)=a\frac{\delta_m'(a)}{\delta_m(1)}\sigma_{8,0}\;.
\end{equation}
Equation \eqref{eq:fs8-parameter} will be our key quantity, which will be tested with the most recent data available in the following sections.

Alternatively, $f\sigma_8(a)$ can also be written as \cite{Ade:2015rim}
\be
f\sigma_8(a)\equiv \frac{\sigma_8^{(vd)}(a)^2}{\sigma_8^{(dd)}(a)},\label{eq:otherfs8}
\ee
where $\sigma_8^{(dd)}(a)$ is the usual $\sigma_8(a)$ parameter as defined above and $\sigma_8^{(vd)}(a)$ is the smoothed density-velocity correlation defined in a similar manner, but using instead the correlation power spectrum $P_{vd}(k)$ and $v=-\nabla v_N/H$ where $v_N$ is the peculiar velocity of the baryons and dark
matter in the Newtonian-gauge, while $d$ is the total matter density perturbation.
Using linear theory for models close to $\Lambda$CDM, it is easy to show that $v=-\nabla v_N/H =-i k \frac{v_N}{aH}= f(a) \delta(k,a)$, see Section 9.2 in Ref.~\cite{Dodelson:2003ft} for a quick derivation. However, the growth can also be written as $\delta(k,a)=\delta_k(k) \; G(a)$, where $\delta_k$ is an initial condition determined from inflation and $G(a)\equiv \frac{\delta(a)}{\delta(1)}$ is the normalized growth, see Eq.~(7.8) in Ref.~\cite{Dodelson:2003ft}. Then, using the definitions of the $\sigma$ parameters, we have
\bea
\sigma_8^{(vd)}(a)^2 &=& \langle v(x,a) \delta(x,a) \rangle \nn \\
&=& f(a) G(a)^2 \langle \delta(x)^2 \rangle\nn \\
&=& f(a) G(a)^2 \sigma_{8,0}^2,
\eea
and
\bea
\sigma_8^{(dd)}(a)^2 &=& \langle \delta(x,a) \delta(x,a) \rangle \nn \\
&=& G(a)^2 \langle \delta(x)^2 \rangle \nn \\
&=& G(a)^2 \sigma_{8,0}^2.
\eea
Finally, using the definition of Ref.~\cite{Ade:2015rim} we have
\bea
f\sigma_8(a)&\equiv& \frac{\sigma_8^{(vd)}(a)^2}{\sigma_8^{(dd)}(a)}\nn\\
&=&\frac{f(a) G(a)^2 \sigma_{8,0}^2}{G(a) \sigma_{8,0}} \nn\\
&=& f(a) G(a) \sigma_{8,0},
\eea
which exactly agrees with our original definition of Eq.~(\ref{eq:fs8-parameter}).

\section{Formalism}\label{formalism}
Here we report the basic equations that will be used to perform our analysis and we refer to \cite{Amendola:2012wc} for the details on the derivation of the internal robustness.
The statistical definition of the Bayesian evidence is
\be
\EE({\bf x}|\,M) =\int L({\bf x}|\,\bm \theta^{M})\pi(\bm \theta^M)\rmd \bm \theta^M,
\ee
where ${\bf x} = (x_1,\,x_2,\,...,\,x_N)$ are the $N$ random data, $\bm \theta^M = (\theta_1,\,\theta_2,\,...,\,\theta_n)$ are the $n$ theoretical parameters of the model $M$. The likelihood function is $L({\bf x}|\,\bm \theta^{M})$, while the prior probability of the parameters of the model is $\pi(\bm \theta^M)$. By the help of the Bayes' theorem we can find the posterior probability $\PP(\,M|\,{\bf x})$ of having a model $M$ given the data
\be
\PP(\,M|\,{\bf x})  = \EE({\bf x}|\,M)\frac{\pi(M)}{\pi({\bf x})}\,.
\ee
By using the last equation, we can compare two different models by considering the ratio of their probabilities:
\be
\frac{\PP(\,M_1|\,{\bf x})}{\PP(\,M_2|\,{\bf x})} = \BB_{12}\frac{\pi(M_1)}{\pi(M_2)}\,;
\ee
with the Bayes factor defined as
\be
\BB_{12} = \frac{\EE({\bf x}|\,M_1)}{\EE({\bf x}|\,M_2)}\,.
\ee
If we assume that the prior probabilities of having two different models are the same, then the Bayes factor alone will help us to favor or disfavor a particular model. If $\BB_{12}>1$ then the data favors the model $M_1$, if it is less than 1, then $M_2$ is favored.


However, the robustness test needs a further assumption, that is: the data have to come from two different distributions. The reason is two fold: first the total evidence can be factorized as the product of the two evidences and, second, which is the underlying meaning of the robustness test, we would like to prove that data are reliable. If the data is partitioned into two subsets, say $\{{\bf x_1},{\bf x_2}\}$ and we assume that they come from two models, say $M_1, M_2$, then the Bayes factor becomes
\be
\BB_{12} = \frac{\EE({\bf x}|\,M_1)}{\EE({\bf x_1}|\,M_1)\EE({\bf x_2}|\,M_2)}\;.
\ee
Finally, we can define the internal robustness as
\be
\text{iR}_{12}=\log{\BB_{12}} = \log\left(\frac{\EE({\bf x}|\,M_1)}{\EE({\bf x_1}|\,M_1)\EE({\bf x_2}|\,M_2)}\right)\;.
\ee
This approach allows us to detect if a subset of the data follows another cosmological model or if a specific survey is affected by systematics and hence altering the measurement itself.

However, the assumption of having two different models is not strictly mandatory and we will choose the same cosmological model for both subsets. In this work we will set the cosmological model to be $\Lambda$CDM and the parameters for both subsets to $\bm \theta = (\Omega_{m,0},\sigma_{8,0})$. Hereinafter, we drop the $M$ superscript and the $1,2$ subscripts, since we only consider one cosmological model.

Our analysis invokes priors on the parameters, for that we choose a flat prior in the $[0,1]$ range for $\Omega_{m,0}$ in order to allow for all physical values possible for the matter density.
On the other hand, the choice of a prior for $\sigma_{8,0}$ is less evident;  since the prior directly affects the evidence value, so we are going to consider three priors for $\sigma_{8,0}$, to assess the impact of the prior selection on the internal robustness.
We choose the following three cases:
\begin{itemize}
	\item \textbf{Narrow flat prior}: this is a typical flat prior in the range $[0.3,1.5]$.
	\item \textbf{Broad flat prior}:  this is a flat prior in the range $[0,10]$, which allows for high values of $\sigma_{8,0}$.
	\item \textbf{Gaussian prior}: the third prior to consider is a Gaussian distribution centered on $0.8150$, with a standard deviation of $0.0087$, based on the Planck 2015 results (TT,TE,EE+lowP+lensing \cite{Ade:2015xua}).
\end{itemize}
It is clear that we only allow for positive values of $\sigma_{8,0}$ in order to remain physical.

The data considered are $f\sigma_8(z)$ measurements (with $z=-1+1/a$ being the redshift of the measurement), and we assume a Gaussian likelihood for the data with a covariance matrix $\bf C$.
We represent the observed data in different redshifts as $ \mathbf{m}=(m(z_1),\dots, m(z_n))$ and its theoretical prediction as $\bm \mu(\bm \theta)=(\mu(z_1),\dots, \mu(z_n))$, which depends on the cosmological model and parameters.
We also take into account the redshift correction of \cite{Nesseris:2017vor}, which features a correction factor of
\begin{equation}
\text{fac}(z^i)= \frac{H(z^i)\,d_A(z^i)}{H^{\text{ref},i}(z^i) \, d_A^{\text{ref},i}(z^i)}\;.
\end{equation}
with the label $\text{ref},i$ stating that the cosmology considered is the reference cosmology used on the corresponding data point on redshift $z^i$. Hence, the corrected theoretical prediction is \cite{Macaulay:2013swa}
\begin{equation}
\mu_c^i=\frac{\mu^i}{\text{fac}(z^i)}\;.
\end{equation}
We are now in the position to define the data vector with the corresponding modification:
\begin{equation}
     \mathbf{x} =  \mathbf{m} - \bm\mu_c.
\end{equation}
Then, the chi-squared is
\begin{equation}
	\chi^2=  \mathbf{x}^T  \mathbf{C}^{-1}  \mathbf{x}\;,
\end{equation}
which is related to the likelihood via $L=e^{-\chi^2/2}/\sqrt{(2\pi)^n |\mathbf{C}|}$.

To speed up the computations, we note that the $\sigma_{8,0}$ parameter can be marginalized theoretically \cite{Basilakos:2016nyg,Taddei:2016iku}. We rewrite the $\chi^2$:
\begin{equation}
\chi^2=  \mathbf{m}^T \mathbf {C}^{-1} \mathbf{m}  -2  \mathbf{m}^T  \mathbf C^{-1} \bm \mu_c + \bm \mu_c^T \mathbf{C}^{-1} \bm \mu_c \;.
\end{equation}
The corrected theoretical prediction marginalized over $\sigma_{8,0}$ will be $\bm \nu =\bm \mu_c / \sigma_{8,0}$. Then, the  $\chi^2$ can be rewritten as
\begin{equation}
\chi^2=\xi_{mm}-2\xi_{m\nu}\sigma_{8,0}+\xi_{\nu\nu}\sigma_{8,0}^2\;,
\end{equation}
where the single terms are:
\begin{align}
\begin{split}
\xi_{mm}=  \mathbf{m}^T   \mathbf{C}^{-1}  \mathbf{m}, \\
\xi_{m\nu}=   \mathbf{m}^T   \mathbf{C}^{-1} \bm \nu, \\
\xi_{\nu\nu}=\bm \nu^T   \mathbf{C}^{-1} \bm \nu.
\end{split}
\end{align}
The posterior probability distribution marginalized over $\sigma_{8,0}$ is
\begin{equation}
\PP(\Omega_{m,0})=\int L(\Omega_{m,0},\sigma_{8,0})\pi(\Omega_{m,0},\sigma_{8,0}) \d\sigma_{8,0}\;.
\end{equation}
We now consider two cases for the prior probability on $\sigma_{8,0}$: a flat prior between $[a,b]$ and a Gaussian prior with mean $s$ and variance $\epsilon^2$.
Let us start by considering the flat prior case. The integration of the posterior is:
\begin{multline}
\PP_f(\Omega_{m,0})=\frac{1}{(b-a)\sqrt{(2\pi)^n |\mathbf{C}|}} \int\limits_a^b  e^{-\chi^2/2}\d\sigma_{8,0}\\
=\frac{1}{(b-a)\sqrt{(2\pi)^n |\mathbf{C}|}} \exp\left(-\frac{1}{2}\left[\xi_{mm}-\frac{\xi_{m\nu}^2}{\xi_{\nu\nu}} \right] \right)  I_f\;,
\end{multline}
where the quantity $I_f$ is:
\begin{multline}
I_f=\int\limits_a^b e^{-\frac{\xi_{\nu\nu}}{2}(\xi_{m\nu}/\xi_{\nu\nu} - \sigma_8)^2} \d\sigma_8 \\
 = \sqrt{\frac{\pi}{2\xi_{\nu\nu}}} \textrm{erf}\left( \frac{\xi_{m\nu} - x \xi_{\nu\nu}}{\sqrt{2\xi_{\nu\nu}}} \right) \bigg\rvert_a^b \;.
\end{multline}
For the Gaussian prior case we find, by discarding negative values:
\begin{multline}
\pi_g(\sigma_{8,0})=\frac{e^{-\frac{1}{2}(s-\sigma_{8,0})^2/\epsilon^2}}{\int_0^\infty e^{-\frac{1}{2}(s-\sigma_{8,0})^2/\epsilon^2}\d\sigma_{8,0}} \\
=\frac{e^{-\frac{1}{2}(s-\sigma_{8,0})^2/\epsilon^2}}{\sqrt{\pi\epsilon^2/2}\, [1+\textrm{erf}(s/\sqrt{2\epsilon^2}))]}\\
=A_g \, e^{-\frac{1}{2}(s-\sigma_{8,0})^2/\epsilon^2}\;,
\end{multline}
where we implicitly defined the normalization constant $A_g$.

The posterior probability distribution function then reads:
\begin{multline}
\PP_g(\Omega_{m,0})=\frac{A_g}{\sqrt{(2\pi)^n |\mathbf{C}|}} \int\limits_0^\infty  e^{-[\chi^2+(s-\sigma_{8,0})^2/\epsilon^2]/2}\d\sigma_{8,0}\\
=\frac{A_g}{\sqrt{(2\pi)^n |\mathbf{C}|}} \exp\left(-\frac{1}{2}\left[\xi_{mm}-\frac{\xi_{m\nu}^2}{\xi_{\nu\nu}} \right] \right)  I_g\;,
\end{multline}
where $I_g$ is equal to:
\begin{multline}
I_g=\int\limits_0^\infty e^{-\frac{\xi_{\nu\nu}}{2}(\xi_{m\nu}/\xi_{\nu\nu} - \sigma_8)^2} e^{-\frac{1}{2}(s-\sigma_{8,0})^2/\epsilon^2} \d\sigma_8 \\
=\sqrt{\frac{\pi \epsilon^2}{2(\epsilon^2 \xi_{\nu\nu}+1)}} \exp\left(-\frac{(\xi_{m\nu}-\xi_{\nu\nu}s)^2}{2\xi_{\nu\nu}(\epsilon^2\xi_{\nu\nu}+1)}\right) \\
\left[1+\textrm{erf}\left(\frac{\xi_{m\nu}\epsilon^2+s}{\epsilon\sqrt{2(\epsilon^2 \xi_{\nu\nu}+1)}}\right)\right],
\end{multline}
which is the multiplication of the exponentials of two Gaussians, which is also the exponential part of a Gaussian distribution.

\section{Data Considerations}\label{data}

\subsection{The data set}
The growth rate data set is based on the Gold-2017 compilation from \cite{Nesseris:2017vor}, consisting of 18 independent measurements of $f\sigma_8(z)$, obtained from redshift space distortion measurements from a variety of surveys. Among these surveys, it is important to note that the three WiggleZ \cite{Blake:2012pj} measurements are correlated, and their covariance matrix is
\begin{equation}\label{WiggleZCov}
      \mathbf{C}_{\text{WiggleZ}}= 10^{-3}
    \left(
         \begin{array}{ccc}
           6.400 & 2.570 & 0.000 \\
           2.570 & 3.969 & 2.540 \\
           0.000 & 2.540 & 5.184 \\
         \end{array}
       \right).
\end{equation}
In addition to the Gold-2017 compilation, we update it with 4 recent measurements from \cite{Zhao:2018jxv}. These points have a covariance matrix given by
\begin{equation}\label{SDSS4Cov}
     \mathbf{C}_{\text{SDSS-IV}}= 10^{-2}
    \left(
         \begin{array}{cccc}
   3.098 & 0.892 &  0.329 & -0.021\\
       0.892 & 0.980 & 0.436 & 0.076\\
       0.329 & 0.436 &  0.490   & 0.350 \\
       -0.021 & 0.076 & 0.350 & 1.124
       \end{array}
       \right).
\end{equation}
\begin{table}[htp]
	\caption[]{Compilation of the $f\sigma_8(z)$ measurements used in this analysis along with the reference matter density parameter $\Omega_{m_0}$ (needed for the growth correction) and related references.}
	\begin{center}
		\begin{tabular}{ccccccccc}
			\hline
			\hline
			$z$     & $f\sigma_8(z)$ & $\sigma_{f\sigma_8}(z)$  & $\Omega_{m,0}^\text{ref}$ & Ref. \\ \hline
			0.02    & 0.428 & 0.0465  & 0.3 & \cite{Huterer:2016uyq}   \\
			0.02    & 0.398 & 0.065   & 0.3 & \cite{Turnbull:2011ty},\cite{Hudson:2012gt} \\
			0.02    & 0.314 & 0.048   & 0.266 & \cite{Davis:2010sw},\cite{Hudson:2012gt}  \\
			0.10    & 0.370 & 0.130   & 0.3 & \cite{Feix:2015dla}  \\
			0.15    & 0.490 & 0.145   & 0.31 & \cite{Howlett:2014opa}  \\
			0.17    & 0.510 & 0.060   & 0.3 & \cite{Song:2008qt}  \\
			0.18    & 0.360 & 0.090   & 0.27 & \cite{Blake:2013nif} \\
			0.38    & 0.440 & 0.060   & 0.27 & \cite{Blake:2013nif} \\
			0.25    & 0.3512 & 0.0583 & 0.25 & \cite{Samushia:2011cs} \\
			0.37    & 0.4602 & 0.0378 & 0.25 & \cite{Samushia:2011cs} \\
			0.32    & 0.384 & 0.095  & 0.274 & \cite{Sanchez:2013tga}   \\
			0.59    & 0.488  & 0.060 & 0.307115 & \cite{Chuang:2013wga} \\
			0.44    & 0.413  & 0.080 & 0.27 & \cite{Blake:2012pj} \\
			0.60    & 0.390  & 0.063 & 0.27 & \cite{Blake:2012pj} \\
			0.73    & 0.437  & 0.072 & 0.27 & \cite{Blake:2012pj} \\
			0.60    & 0.550  & 0.120 & 0.3 & \cite{Pezzotta:2016gbo} \\
			0.86    & 0.400  & 0.110 & 0.3 & \cite{Pezzotta:2016gbo} \\
			1.40    & 0.482  & 0.116 & 0.27 & \cite{Okumura:2015lvp} \\
			0.978   & 0.379  & 0.176 & 0.31 & \cite{Zhao:2018jxv} \\
			1.23    & 0.385  & 0.099 & 0.31 & \cite{Zhao:2018jxv} \\
			1.526   & 0.342  & 0.070 & 0.31 & \cite{Zhao:2018jxv} \\
			1.944   & 0.364  & 0.106 & 0.31 & \cite{Zhao:2018jxv} \\
			\hline
			\hline
		\end{tabular}
	\end{center}
	\label{fs8tab}
\end{table}
Our final data set will be constituted of  $N=22$ data points, shown in Table \ref{fs8tab}, the possible combinations of subsets from them is $2^{N-1}-1=2097151$, and we analyze all of the subsets\footnote{Note that we do not count the combination of the full data set with the empty set $\varnothing$.}. The analysis is possible thanks primarily to the marginalization over $\sigma_{8,0}$, as shown in Sec.~\ref{formalism}.

We do not use the data set of Ref.~\cite{Nesseris:2007pa}, even though some data points are at high redshifts, as these are measurements of the growth-rate $f(a)$, which is affected by the bias $b$, and not of the combination $f\sigma_8$ which has been shown to be bias free \cite{Song:2008qt}.

\begin{figure*}[!t]
	\centering
	\includegraphics[width=\textwidth]{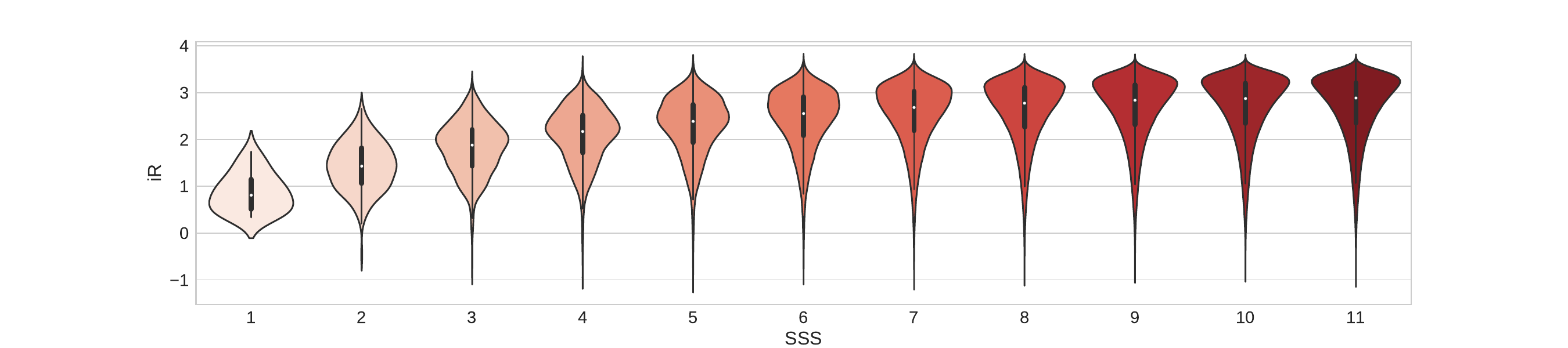}
	\includegraphics[width=\textwidth]{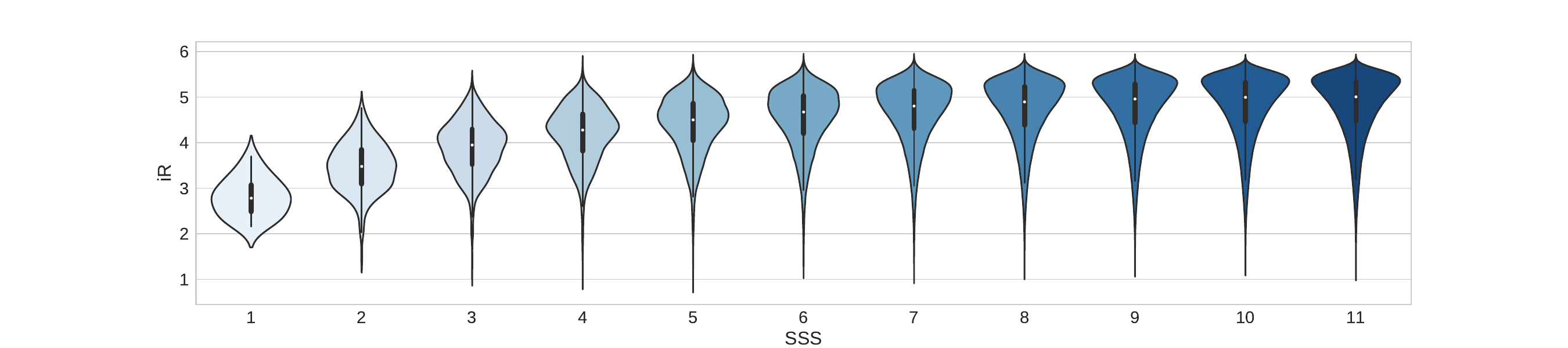}
	\includegraphics[width=\textwidth]{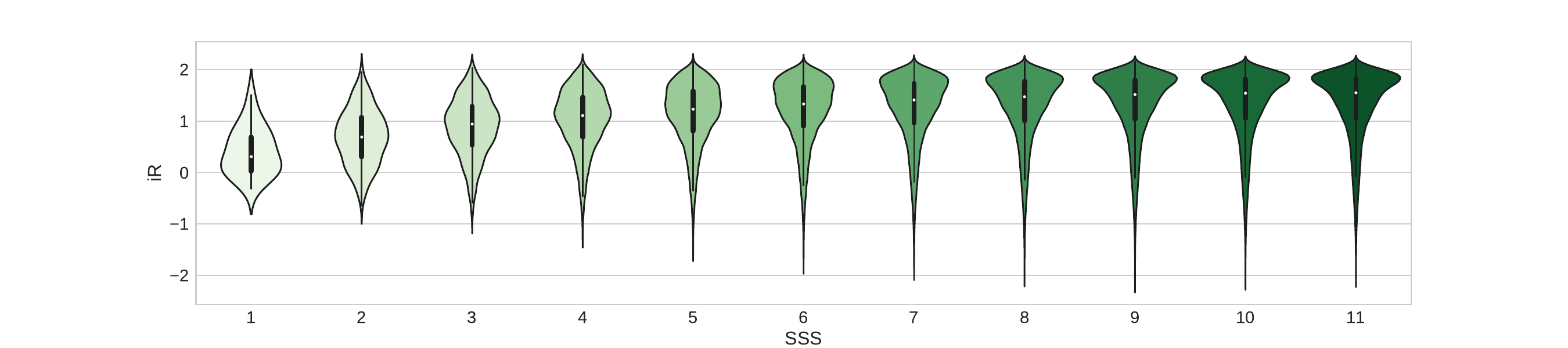}
	\caption[]{Violin plots of the internal robustness distributions, for each of the smaller subset sizes of each partition. We employed the narrow flat prior (upper panel), broad flat prior (medium panel) and Gaussian prior (lower panel). The white dots are the mean value of the internal robustness, the bold black line is the 1$\sigma$ region and the thinner black line is the 2$\sigma$ region.}
	\label{fig:results1}
\end{figure*}

\subsection{Mock Data}
An important feature of this work is the comparison of confidence regions for the probability distributions of the internal robustness (iR-PDF). To obtain the confidence regions, we generate mock data sets based on the form
\begin{equation}
	 f\sigma_8^\text{mock}(z_i)=f\sigma_8(z_i|\;\bm\theta^\text{bestfit})+\mathcal{N}^\text{random} \;,
\end{equation}
meaning that the mock growth rate data is generated from the best fit parameters $\bm\theta^\text{bestfit}$, which are obtained using the complete data set and minimizing the posterior probability (which takes the prior into account). The $\mathcal{N}^\text{random}$ term is evaluated by assuming a Gaussian noise with zero mean and standard deviation equal to those given by the data $\sigma_{f\sigma_8}(z_i)$.

The main reason of comparing the results obtained by using the data and the mock catalogues is to compare directly the iR-PDF to the confidence regions. If the iR-PDF from the data falls off the confidence regions, then we can state that either the data set is not internally robust, meaning that the data set could be affected by systematics, or that a better physical model is required in order to better describe the data. In other words, the mock data confidence regions portrait acceptable offset levels from the best fit cosmology obtained from the complete data set.

For each choice of the prior on $\sigma_{8,0}$, we generate 1000 mock data sets. Then, we sample each one of these data sets into 14000 subset combinations, distributed as follows: 2000 samples for subsets in which the smaller subset size (hereinafter SSS) is 11, another 2000 for subsets with SSS=10, and so on, until SSS=4. We stop at SSS=4 because the number of samples would be larger than the available combinations. As mentioned, the goal is to explore different subset sizes in an equal manner, with the further consideration that, for larger SSS value, we have more possible combinations.

As an ultimate test, we produced mock data sets based on the Planck 2015 best fit parameters \cite{Ade:2015xua}, for which the parameters are $\Omega_{m,0}=0.3121$ and $\sigma_{8,0}=0.815$. The idea behind this is to check whether the tension on  measurements of $\sigma_{8,0}$ between Cosmic Microwave Background (CMB) surveys like Planck and galaxy clustering surveys, see Refs.~\cite{Nesseris:2017vor,Camera:2017tws,Kohlinger:2017sxk,Abbott:2017wau,Gomez-Valent:2017idt,Gomez-Valent:2018nib,Barros:2018efl,Lambiase:2018ows,Gannouji:2018col}, could be due to inconsistencies in the data themselves.

\begin{figure}[!t]
	\centering
	\includegraphics[width=0.5\textwidth]{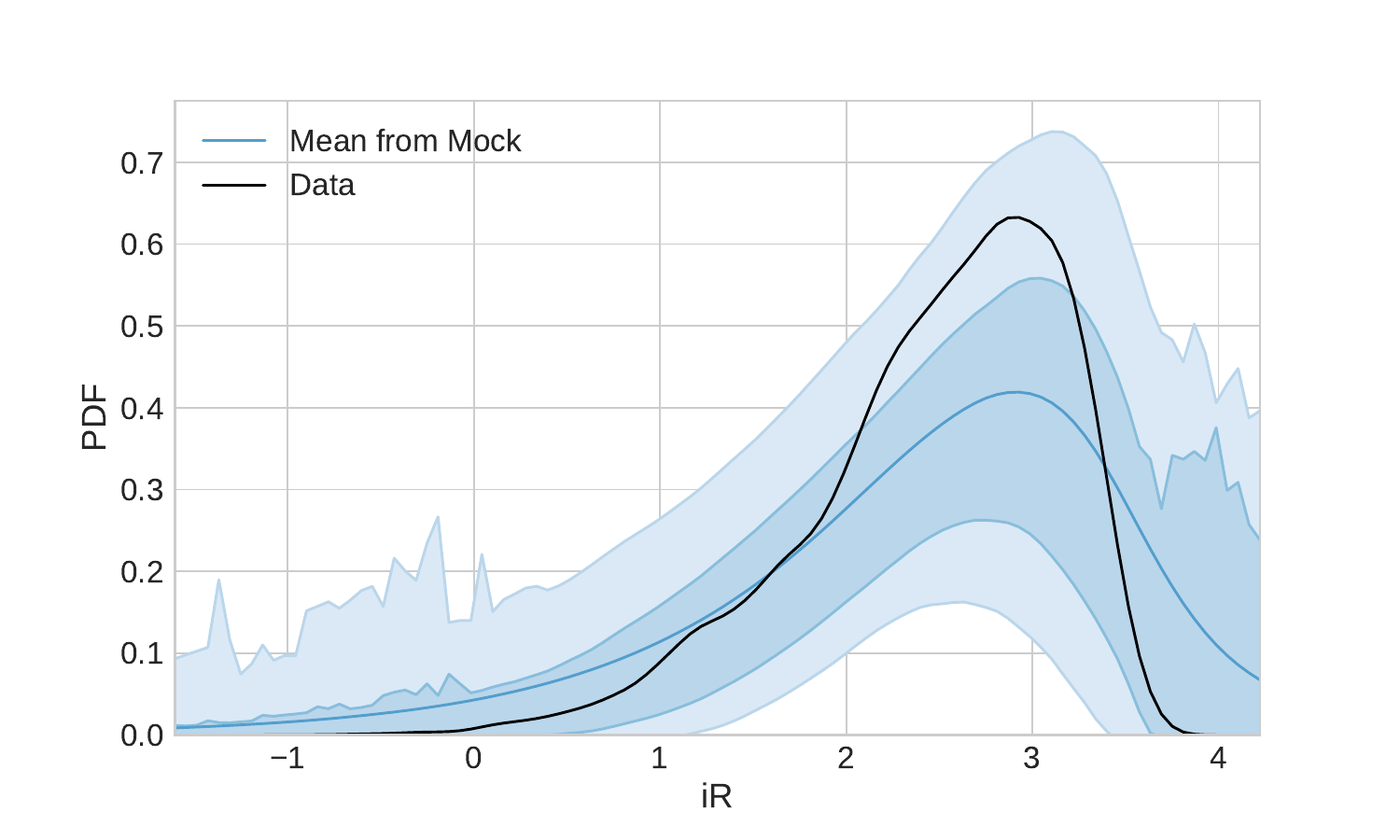}
	\includegraphics[width=0.5\textwidth]{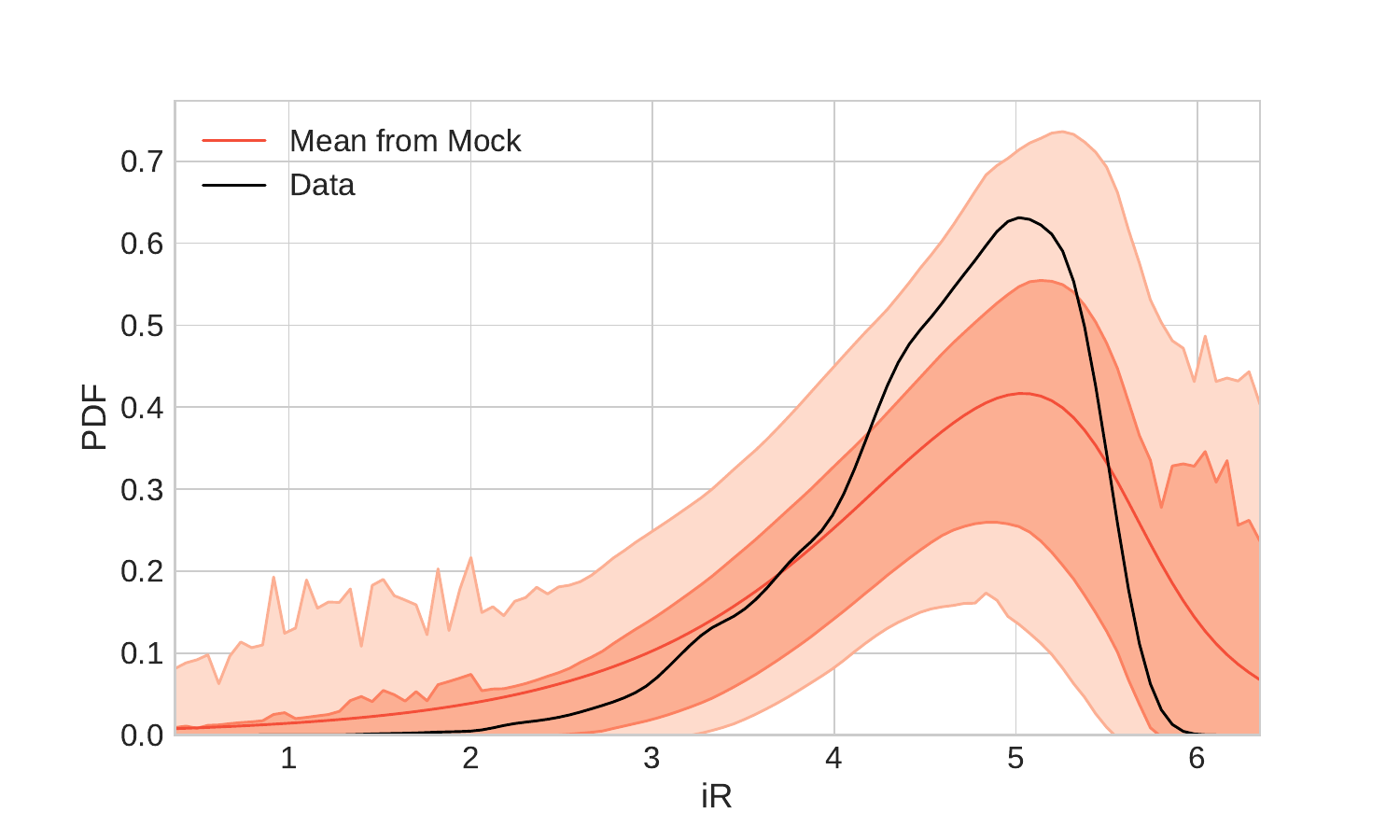}
	\includegraphics[width=0.5\textwidth]{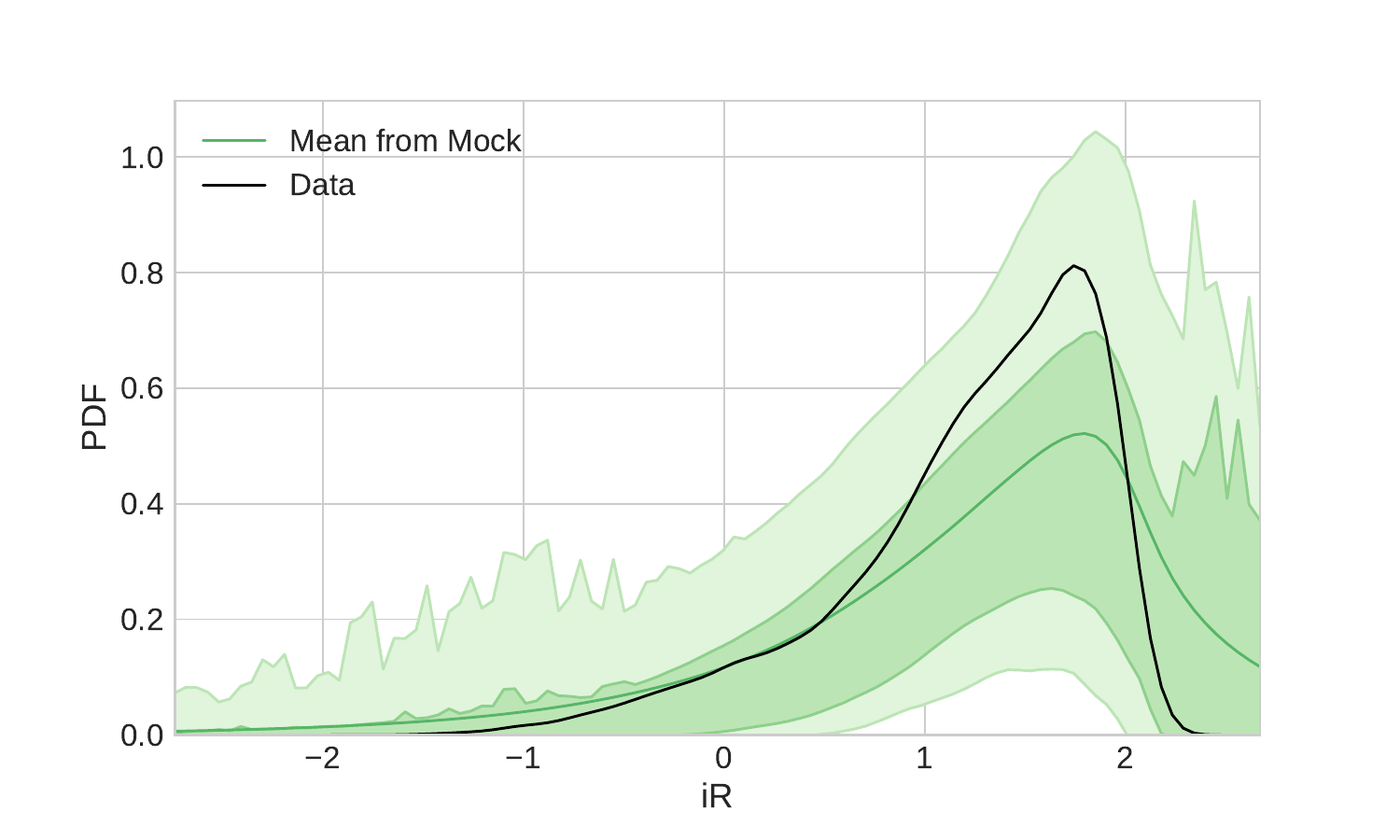}
	\caption[]{Internal Robustness PDF and confidence regions from mock data based on the best fit cosmology, using each prior. They are: narrow flat prior (upper panel), broad flat prior (medium panel) and Gaussian prior (lower panel).}
	\label{fig:results2}
\end{figure}
\begin{figure}[!t]
	\centering
	\includegraphics[width=0.5\textwidth]{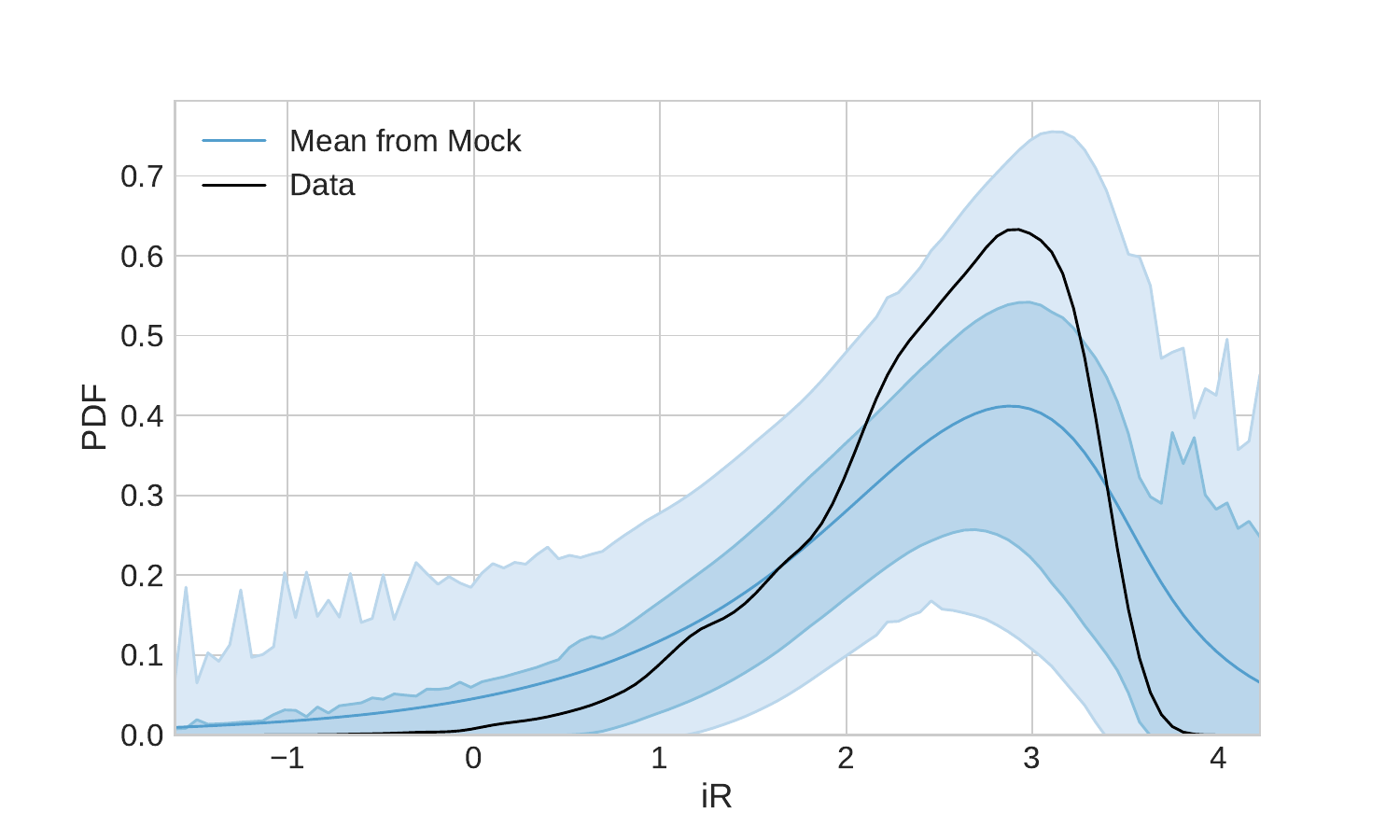}
	\includegraphics[width=0.5\textwidth]{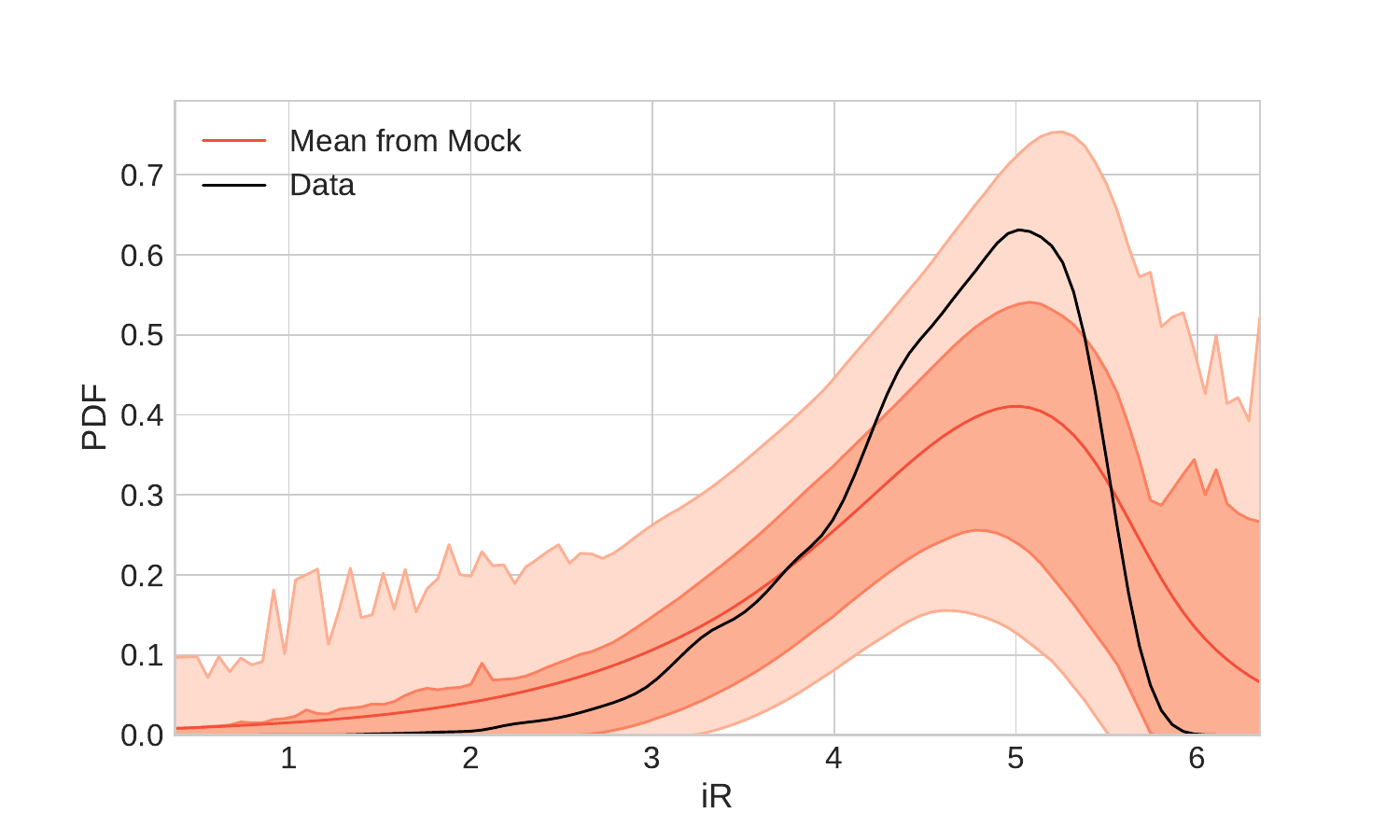}
	\includegraphics[width=0.5\textwidth]{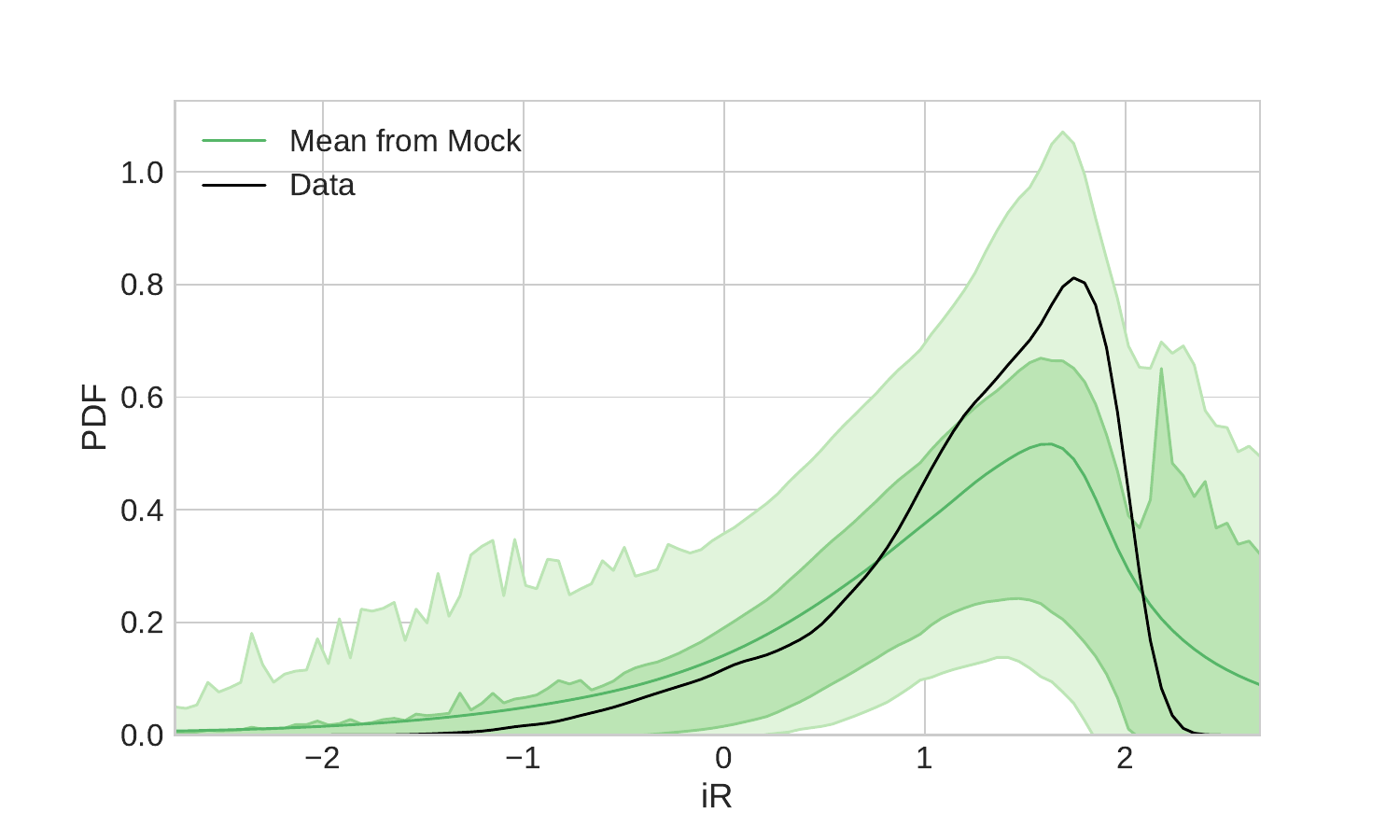}
	\caption[]{Same as Fig. \ref{fig:results2}, but the base parameters for the mock data are from Planck 2015. Priors used: narrow flat prior (upper panel), broad flat prior (medium panel) and Gaussian prior (lower panel).}
	\label{fig:results3}
\end{figure}
\begin{figure*}[!t]
	\centering
	\includegraphics[width=\textwidth]{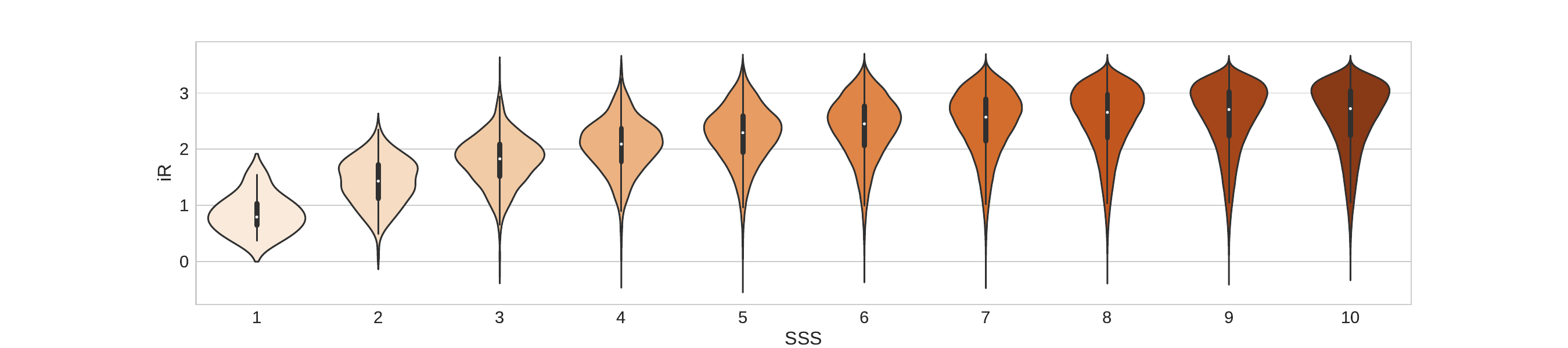}
	\includegraphics[width=\textwidth]{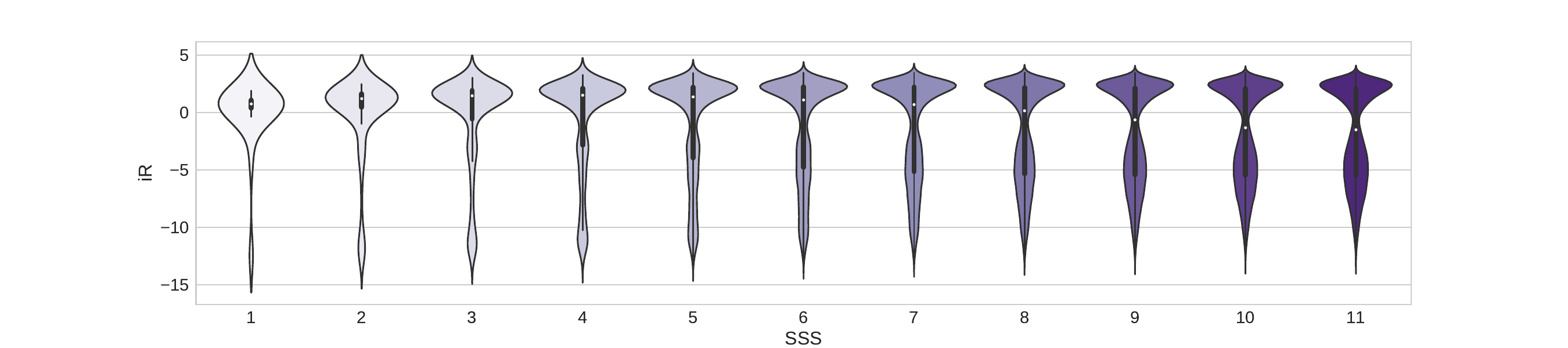}
	\caption[]{Same as Fig. \ref{fig:results1}, but considering the `r26' data set (upper panel) and the `c1' data set (lower panel), both with a narrow flat prior.}
	\label{fig:results4}
\end{figure*}

\begin{figure}[!t]
	\centering
	\includegraphics[width=0.5\textwidth]{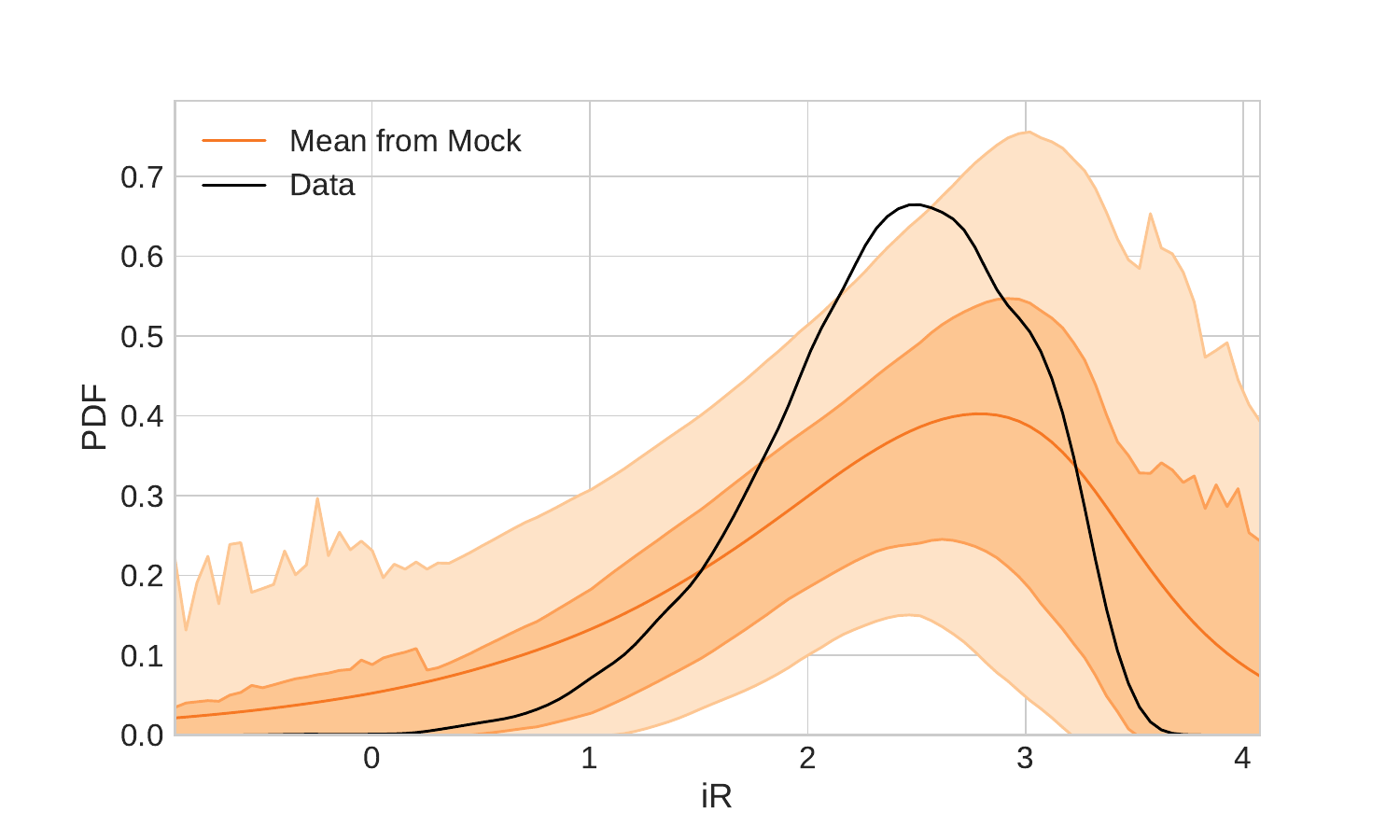}
	\includegraphics[width=0.5\textwidth]{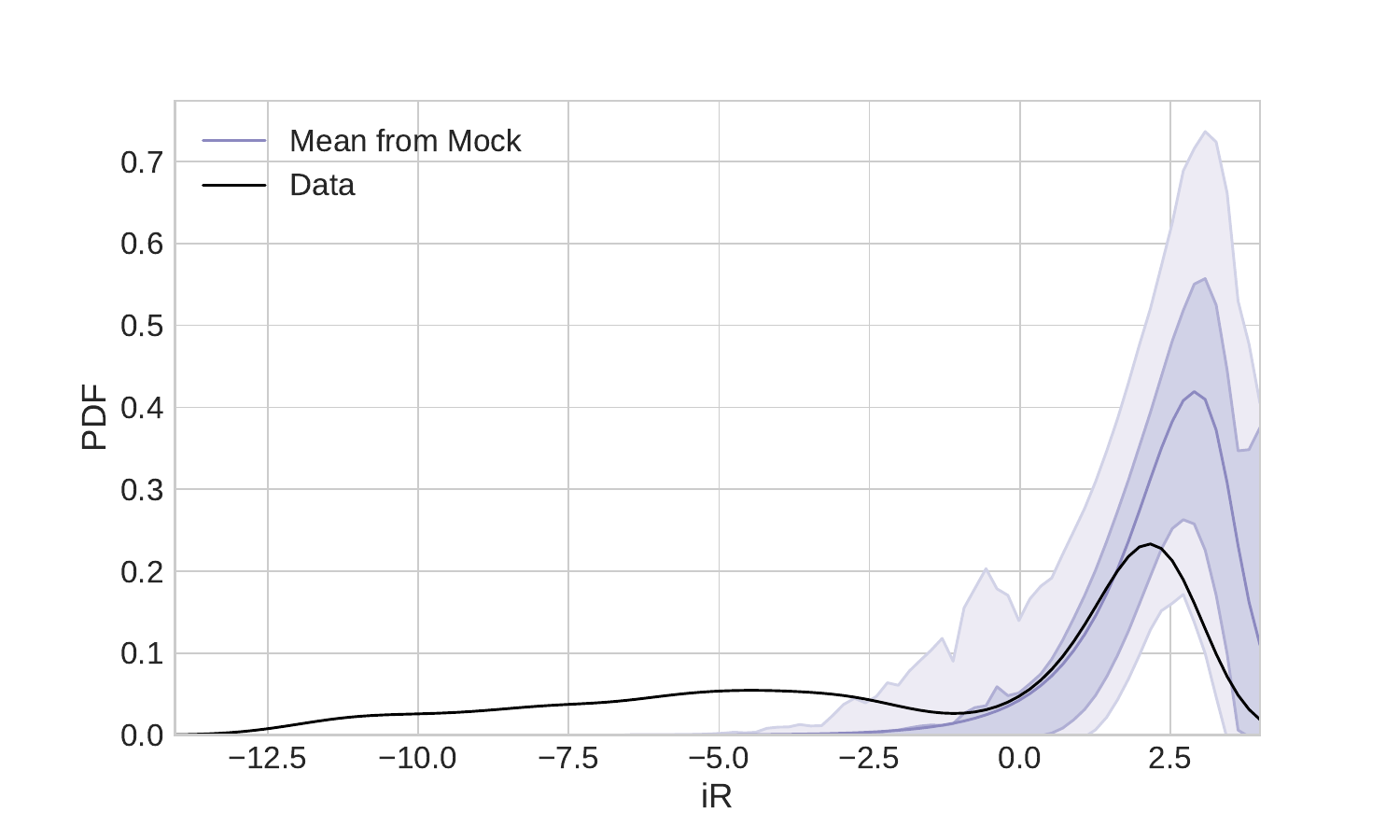}
	\caption[]{Same as Fig. \ref{fig:results2},  but considering the `r26' data set (upper panel) and the `s1' data set (lower panel), both with a narrow flat prior. The mock data used to generate the confidence regions from the upper panel come from the best fit of the `r26' data set, while the ones for the lower panel are from the original data set. }
	\label{fig:results5}
\end{figure}

\subsection{Cross-Checks}

In order to ensure that the method is stable and sensitive to the data set, we decided to opt for two extra cross-checks on our analysis. In brief, the cross-checks have been done using the narrow flat prior only, where we expect the method to be more sensitive to the final results. The cross-checks are:
\begin{itemize}
	\item \textbf{Data removal}: we select one of the combinations with lowest internal robustness and SSS. Then, we remove the data points corresponding to the smaller subset, and evaluate again the complete internal robustness analysis with the new data set. Clearly this procedure forces us to generate a new mock data set with its own best fit. The SSS value will now range from SSS=4 to the maximum SSS possible. In order to be consistent with the number of points, each SSS will be constituted of 2000 sample subsets.
	\item \textbf{Data contamination}: we deliberately choose to \emph{contaminate} the first data point of the data set in order to have a worse iR-PDF. This contamination has been implemented by moving the data point by $5 \sigma$ away from its actual value. In other words, the new first point is constructed as
	\begin{equation}
		f\sigma_8^\text{cont}(z_1)=f\sigma_8(z_1)+5\,\sigma_{f\sigma_8}(z_1)\;.
	\end{equation}
	By moving one of the point by 5$\sigma$ away from its actual position, we expect the iR-PDF to be affected and fall off the confidence regions, clearly the mean iR has to decrease.
\end{itemize}

\section{Results and Discussion}\label{results}

The first results are the complete inspection of the data set, comprising all the possible subset combinations. In Fig.~\ref{fig:results1} we show the iR-PDF in the form of violin plots, arranged by the smaller subset size (SSS) of each subset combination. The three figures differ by the prior used.

From Fig.~\ref{fig:results1} we can see that the internal robustness increases with the SSS. This results was somehow expected as a larger data subsets are less prone to manifest outliers, if the data is free of systematic effects. We also see that the broad prior (middle panel in Fig.~\ref{fig:results1}) has much larger iR than the narrow prior (upper panel in Fig.~\ref{fig:results1}). The difference in the iR value is of the order two regardless of the SSS considered. However, the shape of the distributions changes for SSS$<3$. We can also see that for the Gaussian prior, the distributions are more stretched for small SSS (SSS$<3$) and more clumped up for medium and larger SSS.

In Fig.~\ref{fig:results2} we show the confidence regions of the mock data sets, as reported in Sec.~\ref{data}, along with the iR-PDF of the corresponding prior. The data set black lines were obtained from samples that were equal in size in each SSS as the mock data. We observe that, with the 3 types of prior, the data iR-PDF is within the confidence levels obtained in all the ranges of the internal robustness. This validates the data set, discarding systematic contamination and any other irregularities detectable within the iR formalism.

In Fig.~\ref{fig:results3} we have plot the confidence regions with the Planck-based cosmology. We see that the confidence regions are nearly identical to those of Fig.~\ref{fig:results2}, with the exception of the Gaussian prior case, where the iR-PDF gets closer to the $1\sigma$ region with the Planck-based cosmology mock data. We recall that the Gaussian prior was also chosen based on the Planck 2015 results, so this result is not controversial, although it was not automatically expected, unless we consider the prior to be more constraining than the likelihood alone.

\subsection{Cross-checks results}
As mentioned in the previous section, we decided to make a cross-check analysis to ensure that both method and data set gave sensible results. The first check consisted of removing data points from the subset that gave the lowest robustness.
In our analysis we found that the lowest SSS that gave a negative lowest robustness was constituted of 2 points (hence SSS=2) and the data points falling into this subset were the second and sixth data in the table \ref{fs8tab}. We decided to name this subset `r26'.

Our second cross-check was to take the first data point\footnote{There is no particular reason of choosing the first point. Since the data set is statistically robust, we are allowed to take randomly any point on the list} and move 5 $\sigma$s away from its actual position. The new data set is denominated `c1'. In Fig.~\ref{fig:results4} we show the iR-PDFs for the cross-check data sets.  We can see from the figure that, for the `r26' data set, the iR for SSS$>6$  has a higher minimum but the maximum iR is lower, as well as the mean iR values are smaller with respect to the full data set. This is probably due to, when improving a data set by adding robust points, the iR is expected to increase. On the other hand, the improvement in the minimum iR indeed comes from choosing to remove the points with lower iR on the original data set.

For the second cross-check, i.e. the contamination of one data point, we can see immediately that the internal robustness method detects the inconsistency caused by the contaminated data point, by exhibiting a bimodal shape and a decrease of the iR value when we increase the SSS. These are two features that are not proper of a robust data set.

Finally, in Fig.~\ref{fig:results5} we show the confidence regions for the cross-check data sets. We can see that the confidence regions for the removal cross-check do not  fully contain the iR-PDF. The reason is that, by removing some of the data points, the iR for lower SSS is more affected than those with a higher SSS. This is clear if we consider that the effect of dropping data points are more significant for a small data set rather than a large one, assuming they have the similar weights. The anomaly in the iR-PDF for low SSS can be interpreted as the result of an artificial forcing to avoid small iR values.

For the `c1' data set we clearly see that the deviation from the confidence regions from the mock data confirms the efficacy of the methodology presented.

\section{Conclusions}\label{conclusions}
In this paper we implemented the ``Internal Robustness'' of Ref.~\cite{Amendola:2012wc} to the currently available growth-rate data in the form of $f\sigma_8(z)$ shown in Table \ref{fs8tab}, with the aim to examine the data set for systematics and outlier points in a fully automated manner. The ``Internal Robustness'' is a fully Bayesian approach which is not only sensitive to the local minimum like a standard $\chi^2$ comparison, but also to the entire likelihood and can in principle detect the presence of systematics in the data set. The method works by analyzing combinations of subsets in the data set in a Bayesian model comparison way, potentially finding groups of outliers, data affected by systematics or groups that might follow different physics.

The main goal of our approach was to identify systematic-contaminated data-points, which can then be further analyzed and potentially excluded if they cannot be corrected. Furthermore, in order to validate our analysis and assess its sensitivity we also performed several cross-checks, for example by removing some of the data points or by artificially contaminating some points, while we also generated mock data sets in order to estimate confidence regions of the iR.

We found that, in the first case, when removing the two points with the least robustness the iR for SSS$>6$ has a higher minimum but the maximum iR is lower, as well as the mean iR values are smaller with respect to the full data set. In the second case, by adding an artificially contaminated point which was $\sim 5\sigma$ away from its actual value, we found that the internal robustness method indeed detected the inconsistency caused by the contaminated data point.

In conclusion, implementing the iR methodology we found that the $f\sigma_8(z)$ data set, used in our analysis and shown in Table \ref{fs8tab}, is internally robust showing no anomalous behavior, thus ensuring its internal robustness. This is interesting when discussing the tension of the Planck 15 CMB data and the low redshift measurements coming from galaxy surveys, as we can make sure that the discrepancy does not originate from inconsistencies in the data.

\acknowledgments

B.S. acknowledges help from CONICYT. Powered@NLHPC: This research was partially supported by the supercomputing infrastructure of the NLHPC (ECM-02). He also acknowledges the hospitality of the IFT in Madrid.

S.N. acknowledges support from the Research Project FPA2015-68048-03-3P [MINECO-FEDER], the Centro de Excelencia Severo Ochoa Program SEV-2016-0597 and from the Ram\'{o}n y Cajal program through Grant No. RYC-2014-15843.

D.S. acknowledges financial support from the Fondecyt project number 11140496.

\bibliographystyle{unsrt}
\bibliography{refs}

\end{document}